\documentclass[aps,showpacs,prl,preprint]{revtex4}
\usepackage{amsmath}
\usepackage{graphicx}
\usepackage{amsfonts}
\usepackage{amssymb}%
\begin{document}
\title{Quantum-limited mass flow of liquid $^{3}$He}

\author{G. Lambert$^{a}$,  G.Gervais$^{a}$, and W. J. Mullin$^{b}$}
\affiliation{$^{a}$Department of Physics, McGill University, 3600 rue
Universit\'e, Montr\'eal, Qc, Canada\\
$^{b}$Department of Physics, University of Massachusetts, Amherst,
Massachusetts 01003 USA}

\begin{abstract}
We consider theoretically the possibility of observing unusual quantum
fluid behavior in liquid $^{3}$He and solutions of $^{3}$He in
$^{4}$He systems confined to nano-channels. In the case of pure
ballistic flow at very low temperature conductance will be quantized
in units of $2m^{2}/h$.  We show that these steps should be sensitive
to increases in temperature.  We also use a random scattering
matrix simulation to study flow with diffusive wall scattering.
Universal conductance fluctuations analogous to those seen in electron
systems should then be observable.  Finally, we consider the
possibility of the cross-over to a one-dimensional system at
sufficiently low temperature where the system could form a Luttinger
liquid. 
\vspace{0.2 in}


\end{abstract}

\maketitle

\section{Introduction}

At very low temperatures, the thermal motion of the $^{3}$He atoms in
the liquid becomes very small as the de Broglie wavelength becomes
comparable with the distance between the atoms and so quantum effects
become significant enough to dictate the macroscopic properties of the
liquid.  In a recent paper\cite{MulKal}, one of us noted that
degenerate dilute solutions of $^{3}$He in liquid $^{4}$He might be
cooled by flow through an array of nano-channels whose diameters are
comparable to the de Broglie wave length of the channel.  Aligning the
Fermi energy of a container of fermions with the lowest allowed
band in the channel would allow only the hot gases through the
channel, thereby cooling the remaining gas in the container.  The
presence of such separated bands in a channel immediately suggests
further possible unusual behavior.  For fluids confined over such
small scales, one may legitimately ask whether or not the physics of
the system, for example, the flow properties, are still the same as at
larger scale, or if perhaps a breakdown of the fluid mechanics might
occur.  Indeed recently Sato et al\cite {Sato} considered the
feasibility of seeing quantized conductance in dilute solutions when
there was pure ballistic flow.  Recent advances in materials research
have opened up the possibilities to design and engineer single, or
arrays of pores with diameters of tens of nanometer in ceramic
membranes\cite {Commercial}, and down to only $\sim $1 nm in
tailor-made carbon nanotubes membranes\cite{Holt06}.  These materials
provide a radial length scale small enough to quantize the transverse
motion of the helium atoms so that a realization of quantum-limited
ballistic flow of helium, similar to that observed in the electronic
transport of quantum wires\cite{Wees}, truly seems within experimental
reach.

We consider here possible experiments on pure $^{3}$He as well as
dilute solutions.  We first examine the temperature dependence of the
conductance quantization for pure ballistic flow.  We then investigate
the effect of disorder and how wall scattering in diffusive flow
causes the onset of ``universal'' conductance fluctuations in the mass
flow.  In the ultra-low temperature limit $T\rightarrow 0$, we also
argue that one-dimensional quantum fluids should crossover to an
entirely new exotic type of quantum matter known theoretically as a
Luttinger liquid\cite{Luttinger}.  Such liquid should possess entirely
different flow properties and excitation spectra owing to its
`spin-mass' separated ground state\cite{Affleck}.

This study presented here constitutes only a preliminary analysis to
illustrate the richness of flow properties that one might expect in
experiments on helium in nanoscaled cylindrical pores.  More thorough
investigations at the theoretical and experimental level are in
progress.

\section{Quantized Conductance}

The de Broglie wave length $\lambda _{dB}$ gives the scale of the
diameters of the nanochannels necessary for seeing unusual quantum
effects; the Fermi temperature $T_{F}$ tells us what temperature we
need for degeneracy.  For pure $^{3}$He $\lambda _{dB}\approx 0.7$ nm,
$T_{F}\approx 2$ K while for a 1\% solution of $^{3}$He in liquid
$^{4}$He, $\lambda _{dB}\approx 3$ nm and $T_{F}\approx 120$ mK.
Moreover, the interparticle mean free path is in the micron range
because it contains the factor $(T_{F}/T)^{2}.$ In each case a
dilution refrigerator is able to produce adequate degeneracy.
Ref.~\cite {Sato} considered the possibility of using such arrays of
channels to see the quantization of the conductance.  Unfortunately
dispersion in the channel size can smooth out the quantization steps.
In the present study we will ignore this problem and assume that
technology will soon provide the feasibility of detecting flow from a
single channel or a very small number of channels.  Initially we will
neglect any interparticle scattering and many-body interaction; the only scattering 
we will consider is that from the walls.

A natural definition of the mass conductance, analogous to that used in the
case of an electric current, is given by\cite{Sato} 
\begin{equation}
G=\frac{I}{\Delta \mu /m}
\end{equation}
where $I$ is the mass current, $m$ the particle mass, and $\Delta \mu $ the
difference in chemical potential between the ends of the channel. The mass
current is taken to be 
\begin{equation}
I=\frac{2m}{h}\sum_{n}\int_{-\infty}^{\infty }dp_{z}\frac{p_{z}}{m}\left[
n(\epsilon _{z}+\epsilon _{n}-\mu -\Delta \mu )-n(\epsilon _{z}+\epsilon
_{n}-\mu )\right] \mathcal{T}(\epsilon _{z})  \label{Land1}
\end{equation}
where $p_{z}$ is the longitudinal momentum, and we have
written the energy in the channel as a longitudinal part $\epsilon
_{z}=p_{z}^{2}/2m,$ plus $\epsilon _{n},$ the discrete transverse energies
in the channel. The Fermi distribution is 
\begin{equation}
n(\epsilon -\mu )=\frac{1}{e^{\beta (\epsilon -\mu )}+1}.
\end{equation}
The quantity $\mathcal{T}(\epsilon _{z})$ is the transition
probability for the scatterers in the channel, and $\beta =1/k_{b}T$.
In Eq.~(\ref{Land1}) we have assumed the temperature in the channel to
be uniform; any gradient can easily be incorporated into the formula.
Eq.~(\ref{Land1}) is just the Landauer formula\cite{Landauer} for the
conductance.  We have assumed $\mathcal{T}$ for each band is just a
function of $\epsilon _{z}.$ Expanding the Fermi distributions in
small $\Delta \mu $ we find
\begin{equation}
G=\frac{2m^{2}}{h}\sum_{n}\int_{0 }^{\infty }dz\frac{e^{z-\alpha _{n}}%
}{(e^{z-\alpha _{n}}+1)^{2}}\mathcal{T}(k_{b}Tz)  \label{FiniteT}
\end{equation}
where $z=\beta \epsilon _{z}$ and $\alpha _{n}=$ $\beta (\mu -\epsilon
_{n}). $ Note that if $z<0,$ the transmission probability vanishes.

\begin{figure}[tbp]
\begin{center}
\includegraphics[width=3.4861in]{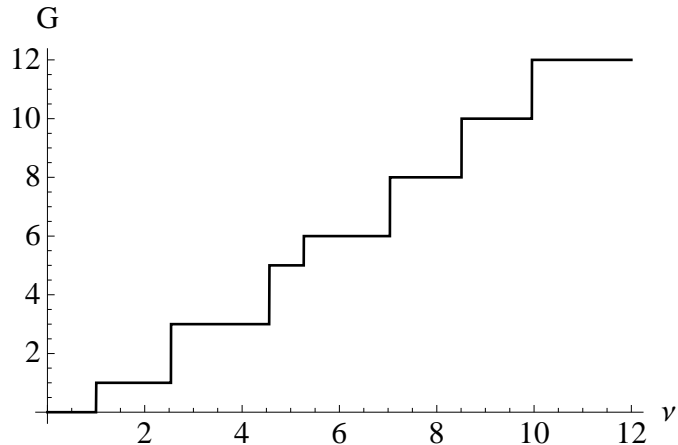}
\end{center}
\caption{Conductance versus $\nu =\epsilon _{F}/\epsilon _{0}$ (where 
$\epsilon _{0}$ is the lowest band edge) for pure ballistic flow at $T=0$.}
\label{G0}
\end{figure}

For extremely low temperature the derivative of the Fermi distribution is
sharply peaked around $z=\alpha _{n}$ so we can write 
\begin{equation}
G=\frac{2m^{2}}{h}\sum_{n}\mathcal{T}(\epsilon _{F}-\epsilon _{n})
\label{ZeroT}
\end{equation}
where $\epsilon _{F}$ is the Fermi energy.  The states in a
cylindrical nano-channel are denoted by $(n,m,k_{z})$ where $n$ and
$m$ are the radial and azimuthal quantum numbers and $k_{z}$ is the
longitudinal wave number.  We assume that the states of different $m$
are all degenerate and that a band is characterized by
$(n,k_{z}^{(n)})$ and a degeneracy factor.  If an atom of energy $E$
enters a channel from a reservoir whose Fermi energy matches a state
in the $N$th band, then it can be in a superposition of bands
$(1,k_{z}^{(1)})\;$to $(N,k_{z}^{(N)})$ such that $E=\epsilon
_{n}+\hbar ^{2}k_{z}^{(n)2}/2m$.  If there is no scattering (that is,
there is only specular scattering at the walls and the latter simply
provide boundary conditions for the wave functions), then particles in
each $k_{z}$ state stay there and do not exchange from one band to
another.  The transmission probability for any band is just a step
function
\begin{equation}
\mathcal{T}(\epsilon _{z})=\Theta (\epsilon _{z}).
\end{equation}
The conductance is then
simply 
\begin{equation}
G=\frac{2m^{2}}{h}N.
\end{equation}
This function is plotted in Fig.~\ref{G0}. The conductance is clearly
quantized in units of 2$m^{2}/h.$ Double steps occur because of a two-fold
degeneracy of some states. 

\begin{figure}[tbp]
\begin{center}
\includegraphics[width=3.4861in]{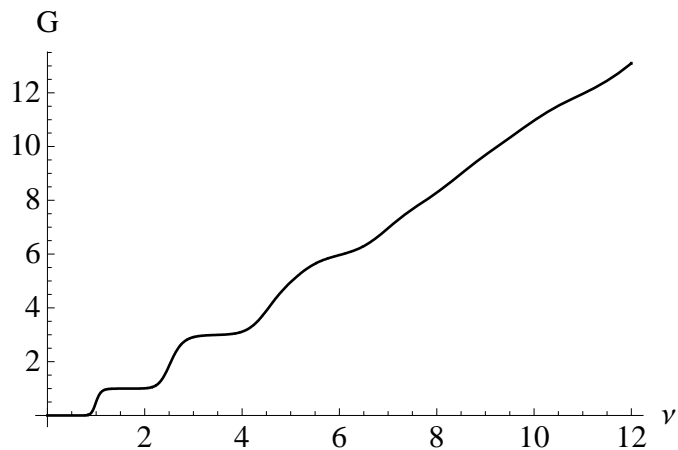}
\end{center}
\caption{Conductance versus $\nu $ at $T/T_{F}=0.05$ for ballistic
flow.}
\label{G05}
\end{figure}

\begin{figure}[tbp]
\begin{center}
\includegraphics[width=3.4861in]{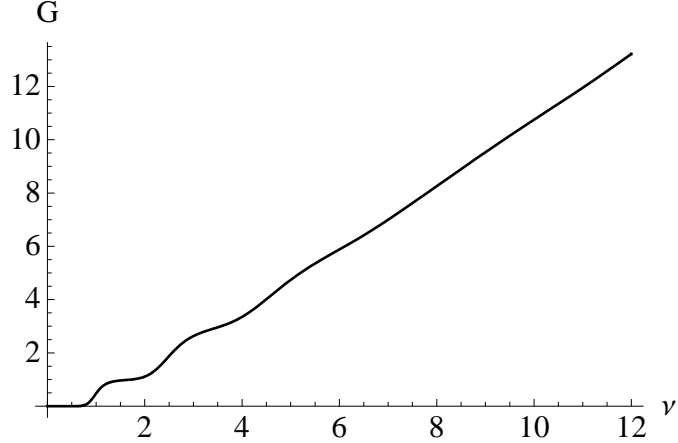}
\end{center}
\caption{Conductance versus $\nu $ at $T/T_{F}=0.1$ for ballistic
flow.}
\label{G1}
\end{figure}

Suppose now we turn up the temperature while still assuming that particles
in the channel maintain their $k_{z}$ values and do not elastically scatter
to other bands. Eq.~(\ref{Land1}) can be written as 
\begin{eqnarray}
G &=&\frac{2m^{2}}{h}\sum_{n}\int_{0 }^{\infty }dz\frac{%
e^{z-\alpha _{n}}}{(e^{z-\alpha _{n}}+1)^{2}}\Theta (z)\nonumber \\
&=&\frac{2m^{2}}{h}\sum_{n}\frac{1}{(e^{-\alpha _{n}}+1)}.
\end{eqnarray}
The states at the Fermi energy feeding into the channel are not sharp and
the steps are rounded as shown in Figs. \ref{G05} and \ref{G1}. The steps
are already reduced substantially at $T/T_{F}=0.1$. The steps are completely
gone by $T/T_{F}=0.5.$ 

\section{Conductance fluctuations}

The above analysis considered only ballistic flow. However wall scattering
can cause diffusive flow. We return here to the $T=0$ situation. An atom
entering the channel is in a superposition of bands $(1,k_{z}^{(1)})\;$to $%
(N,k_{z}^{(N)})$ as above. Backscattering is possible so that the state at
the entrance (left end) to the channel is 
\begin{equation}
\psi _{left}=\left( 
\begin{array}{l}
\phi _{+} \\ 
\phi _{-}
\end{array}
\right) ,
\end{equation}
where $\phi _{i+}=\{(1,k_{z}^{(1)}),\cdots ,(N,k_{z}^{(N)})\}$ and $\phi
_{-}=\{(1,-k_{z}^{(1)}),\cdots ,(N,-k_{z}^{(N)})\}.$ That is, $\psi _{left}$
is a $2N$ dimensional vector. This state has $N$ scattering channels. An
elastic scattering of the particle from $(n,k_{z}^{(n)})$ can be to any of
the $N$ bands as long as energy remains the same. If $\epsilon _{0}<E<$ $%
\epsilon _{1},$ only one channel is available and we have simple 1D
scattering.

The Landauer formula\cite{Landauer} for multichannel conductance is 
\begin{equation}
G=\frac{2m^{2}}{h}\sum_{a,b}T_{ab}
\end{equation}
where $T_{ab}$ is the transmission coefficient for scattering from channel $a$
to channel $b.$

If we assume that a scattering center on the wall has random position and
strength then a method that simply uses a series of completely random
scattering matrices gives a suitable simulation. Such random matrix theory
has been used extensively to treat conductance in electron systems.\cite
{Been}. If the state after a scattering event is

\begin{equation}
\psi ^{\prime }=\left( 
\begin{array}{l}
\phi _{+}^{\prime } \\ 
\phi _{-}^{\prime }
\end{array}
\right).
\end{equation}
Then the transfer matrix $M$ satisfies 
\begin{equation}
\psi ^{\prime }=M\psi
\end{equation}
or 
\begin{equation}
\left( 
\begin{array}{l}
\phi _{+}^{\prime } \\ 
\phi _{-}^{\prime }
\end{array}
\right) =\left( 
\begin{array}{ll}
M_{++} & M_{+-} \\ 
M_{-+} & M_{--}
\end{array}
\right) \left( 
\begin{array}{l}
\phi _{+} \\ 
\phi _{-}
\end{array}
\right)
\end{equation}
where each $M_{ij}$ is an $N\times N$ matrix. The $S$ matrix, also a $%
2N\times 2N$ matrix that we write as 
\begin{equation}
S=\left( 
\begin{array}{ll}
r & t \\ 
t^{\prime } & r^{\prime }
\end{array}
\right),
\end{equation}
connects outgoing (on the left of the equation) and incoming waves (on the
right of the equation) according to 
\begin{equation}
\left( 
\begin{array}{l}
\phi _{-} \\ 
\phi _{+}^{\prime }
\end{array}
\right) =\left( 
\begin{array}{ll}
r & t \\ 
t^{\prime } & r^{\prime }
\end{array}
\right) \left( 
\begin{array}{l}
\phi _{+} \\ 
\phi _{-}^{\prime }
\end{array}
\right).
\end{equation}
where $t$, $t^{\prime}$, $r$, and $t^{\prime}$ are $N\times N$
matrices.  Suppose we have a two-channel system with just one
scattering site so that the final state has $\phi _{-}^{\prime }=0$.  That
is, after the scattering one has only the transmitted wave.  Then we
have $\phi _{-}=r\phi _{+}$ and $\phi _{+}^{\prime }=t\phi _{+};$ $r$
represents reflection and $t$ transmission.  In detail for the
two-channel system we have
\begin{equation}
\left( 
\begin{array}{l}
\phi _{+1}^{\prime } \\ 
\phi _{+2}^{\prime }
\end{array}
\right) =\left( 
\begin{array}{ll}
t_{11} & t_{12} \\ 
t_{21} & t_{22}
\end{array}
\right) \left( 
\begin{array}{l}
\phi _{+1} \\ 
\phi _{+2}
\end{array}
\right).
\end{equation}
If a particle starts in channel 1 ($\phi _{+2}=0)$, the probability of
scattering to final channel 1 is $\left| t_{11}\right| ^{2}$ and the
probability of scattering to final channel 2 is $\left| t_{21}\right| ^{2}.$
The total probability of scattering out of channel 1 is $T_{1}=\left|
t_{11}\right| ^{2}+\left| t_{21}\right| ^{2}.$ The total conductance for an
incoherent mixture of all incoming states is then 
\begin{equation}
G=\frac{2m^{2}}{h}\sum_{a,b}\left| t_{ab}\right| ^{2}=\frac{2m^{2}}{h}%
\mathrm{Tr}[t^{\dagger }t].
\end{equation}
The same result holds for many channels.

The $S$ matrix is unitary and symmetric if we impose current conservation
and time-reversal invariance, which also impose conditions on the $M$
matrix. We want a whole series of randomly chosen transfer matrices because
the transfer matrix for $P$ consecutive scatterings satisfies 
\begin{equation}
M_{tot}=M_{1}M_{2}\cdots M_{P}.
\end{equation}
Because $S$ and $M$ both connect $\phi _{\pm a}$ and $\phi _{\pm a}^{\prime
} $ we can relate the elements of one to the other to find $M$. We then use 
\begin{equation}
t=\left( M_{--}\right) ^{-1}.
\end{equation}

\begin{figure}[tbp]
\begin{center}
\includegraphics[width=3.4861in]{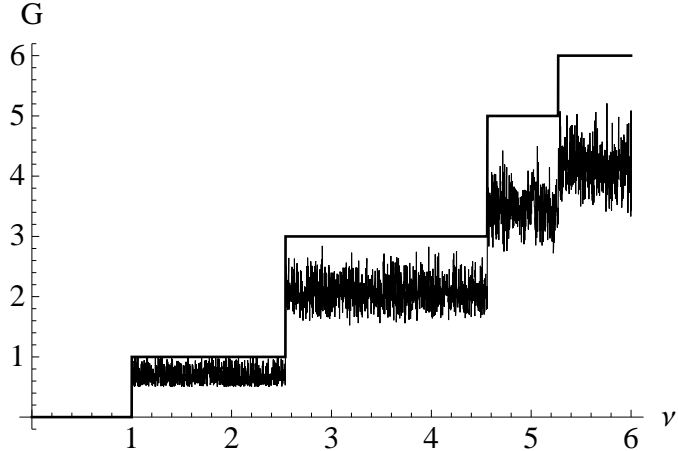}
\end{center}
\caption{Conductance versus $\nu $ for one scatterer. The regular
stepped line is the conductance with pure ballistic flow.}
\label{GR1}
\end{figure}

It is not difficult to construct these random matrices. Some results are
shown in the figures. In Fig.~\ref{GR1}, there is only a single scattering.
Steps are clearly established depending on the number of channels. On the
other hand, there are fluctuations in the conductance having amplitude close
to one conductance unit; these are the ``universal conductance
fluctuations'' observed in electron systems.\cite{UCF} 

\begin{figure}[tbp]
\begin{center}
\includegraphics[width=3.4861in]{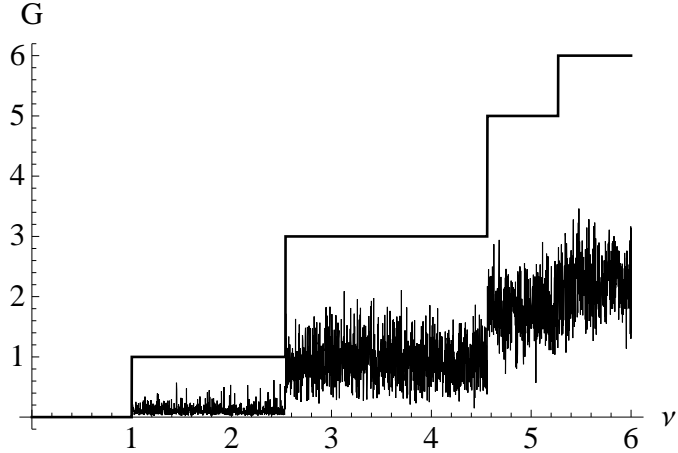}
\end{center}
\caption{Conductance versus $\nu $ for 3 scatterers.}
\label{GR3}
\end{figure}

\begin{figure}[tbp]
\begin{center}
\includegraphics[width=3.4861in]{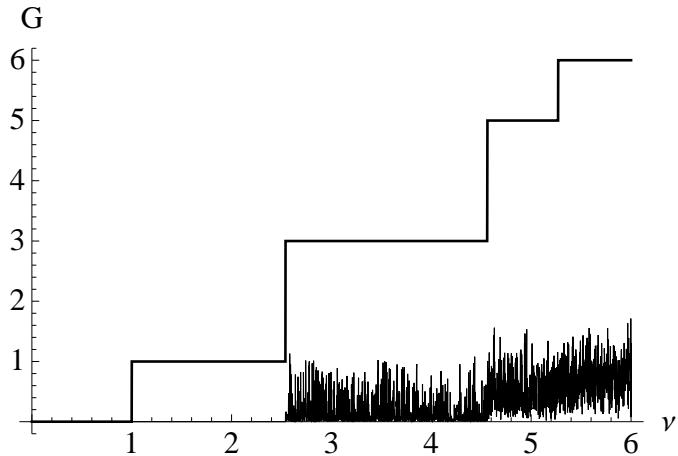}
\end{center}
\caption{Conductance versus $\nu $ for ten scatterers.}
\label{Gr10}
\end{figure}

In Fig.~\ref{GR3} and Fig.~\ref{Gr10} we increase the number of scatterers
to three and ten, respectively. The step sizes are reduced in each case, as
one might expect from an increase in backscattering. All steps maintain the
one unit fluctuation value except the single-channel case, which is strongly
reduced for three scatterers and has disappeared for ten. One-dimensional
motion has reached full localization for 10 scatterers. In that case the
step structure has almost disappeared as well. 

This random matrix method seems to capture many of the effects we expect
experimentally. However this method is so generic that it says nothing
specific, for example, 
about the scattering being just at the walls of the channels.
Alternative techniques more specific to helium are being investigated.

\section{Luttinger liquids?}

Based on simple estimates of the transverse quantum zero-point motion,
we argue that in the regime of nanometric pore size and at
sufficiently low temperatures, the quantum flow properties may not be described
by a fluid with independent particles, as used above, but can enter a new kind
of low-dimensional $^{3}$He ground state with experimentally unknown
excitation spectra and transport properties.  In this regime, the
$^{3}$He Fermi system should reach the strongly interacting
one-dimensional limit for which a breakdown of the Landau Fermi
liquid picture is expected\cite{Luttinger} and the system becomes a
Luttinger liquid.  This expected crossover to the one-dimensional
regime should be considered in a theory describing the transport of
$^{3}$He particles inside very narrow channels. The ideal gas
approximation is no longer valid in this situation and interactions
are vital to the existence of this phase. 

As an idealized case, let's consider the transport of pure $^{3}$He
inside a nanochannel with a diameter $d\sim 1$ nm and of length $L$
such that $L\gg d$.  If the fluid is effectively confined
\textit{radially} in a cylindrical nanochannel, the fluid will be in
the one-dimensional limit when  the occupied states are all in the
lowest band. The degree of occupation is measured by the 
one-dimensional Fermi energy, given by
\begin{equation}
T_{F1D}=\frac{\hbar ^{2}}{2m_{3}k_{B}}\left( \frac{\pi n_{1D}}{2}\right)
^{2},
\end{equation}
where the number density $n_{1D}$ is the reciprocal of the interatomic
distance.  A characteristic temperature measuring the separation
between bands is of order
\begin{equation}
 T_{1D}^{\star }=\frac{\hbar ^{2}}{m_{3}d^{2}k_{B}}.
\end{equation}
We expect a crossover to the one-dimensional regime in nanochannels to
require $T_{F1D}<T_{1D}^{\star }$.

Assuming an interatomic distance of $\sim 1$ nm, we estimate
$T_{F1D} \sim 200$ mK and $T_{1D}^{\star }\sim 160$ mK for $d=1$ nm.
With $T_{F1D}$ and $T_{1D}^{\star }$ comparable, reaching the 1D
liquid in the pure case is a possiblity if the temperature is low
enough.  In addition, a new unidimensional state of matter will arise
only when the $^{3}$He atoms are in the strongly correlated regime
attained at temperature $T\ll T_{F1D}$, say, at $T\lesssim
0.1T_{F1D}\sim 20$ mK. The conditions needed may be within experimental
reach for a few nanometer hole in the pure case.

For a 1\% concentration of isotopic solution the one-dimensional Fermi
temperature is of order 10 mK, and $T_{1D}^{\star }\sim 17$ mK for
$d=3$ nm.  The crossover around $\sim 1$ mK is within the regime of
the cooling technique by filtering discussed by Ref.~\cite{MulKal}.
The use of isotopic helium solutions may allow one to adjust the
parameters into the needed ranges.

The excitation spectrum of the one-dimensional fermionic system differs
entirely from a Fermi liquid, and, for electrons, this state has also
been known to give rise to interesting power law transport behaviour.
Depending on whether the interactions are attractive or repulsive, the
1D-quenched $^{3}$He system could lead to interesting new physics
analogous to that of one-dimensional electronic system such as, for
example, the highly sought-after spin-charge separation for which
there are now important experimental hints\cite{Amir}.  Having both
pure $^{3}$He and dilute solutions available allows the possibility of
a considerable alteration of the interatomic interactions leading to
differing physical properties.

\section{Conclusion}

We have shown that many of the interesting effects known for electron
systems can be studied experimentally and theoretically in $^{3}$He in
nanochannels.  We have provided examples of quantum-limited mass flow
giving rise to a quantization of the conductance.  We have also
calculated its temperature dependence and have shown the appearance of
conductance fluctuations when diffusive scattering takes place.  We
have also argued that a very interesting new state of matter, a
Luttinger liquid, which may have quite new transport properties and
excitations, could perhaps form in $^{3}$He confined to nano-channels.
While this theoretical study has been of a preliminary nature, we
present it in the hopes of stimulating further discussion of this
interesting system.

We would like to acknowledge helpful discussions with I. Affleck.
This work has been supported by the Natural Sciences and Engineering
Research Council of Canada (NSERC), the Canadian Institute for
Advanced Research (CIFAR), and the Alfred P. Sloan Foundation for their
support under their fellowship program (G.G).

\end{document}